**Engineering Vacancies in $Bi_2S_3$ yields sub-Bandgap Photoresponse and highly sensitive Short-Wave Infrared Photodetectors**

*Nengjie Huo, Alberto Figueroba, Yujue Yang, Sotirios Christodoulou, Alexandros Stavrinadis, César Magén, Gerasimos Konstantatos\**

Dr. N. Huo, Dr. A. Figueroba, Dr. Y. Yang, Dr. S. Christodoulou, Dr. A. Stavrinadis, Prof. G. Konstantatos
ICFO – Institut de Ciencies Fotoniques, The Barcelona Institute of Science and Technology, Castelldefels, 08860, Barcelona, Spain
E-mail: Gerasimos.konstantatos@icfo.es

Dr. Y. Yang,
School of Physics and Optoelectronic Engineering, Guangdong University of Technology, Guangzhou 510006, China

Prof. C. Magén,
Instituto de Ciencia de Materiales de Aragón (ICMA), Universidad de Zaragoza-CSIC, 50009 Zaragoza, Spain
Laboratorio de Microscopías Avanzadas, Instituto de Nanociencia de Aragón, Universidad de Zaragoza, 50018 Zaragoza, Spain
Departamento de Física de la Materia Condensada, Universidad de Zaragoza, 50009 Zaragoza, Spain.

Prof. G. Konstantatos
ICREA – Institució Catalana de Recerca i Estudis Avançats, Lluis Companys 23, 08010 Barcelona, Spain

**Abstract:** Defects play an important role in tailoring the optoelectronic properties of materials. Here we demonstrate that sulphur vacancies are able to engineer sub-band photoresponse into the short-wave infrared range due to formation of in-gap states in $Bi_2S_3$ single crystals supported by density functional (DF) calculations. Sulfurization and subsequent refill of the vacancies results in faster response but limits the spectral range to the near infrared as determined by the bandgap of $Bi_2S_3$. A facile chemical treatment is then explored to accelerate the speed of sulphur deficient (SD)-based detectors on the order of 10 ms without sacrificing its spectral coverage into the infrared, while holding a high D* close to $10^{15}$ Jones in the visible-near infrared range and $10^{12}$ Jones at 1.6 µm. This work also provides new insights into the role sulphur vacancies play on the electronic structure and, as a result, into sub-bandgap photoresponse enabling ultrasensitive, fast and broadband photodetectors.

1. Introductoin

High performance photodetectors have hitherto relied on exciting semiconductors with photon energies at least equal to their bandgap. Therefore the selection of a semiconductor material to address a specific spectral band has been a determinant factor. It is well known that silicon covers very efficiently in both performance and cost the visible and near infrared part of spectrum down to ~1100 nm, whereas applications that require photodetection in the short-wave infrared have relied to III-V InGaAs or Ge photodetectors with the associated challenges in their integration onto CMOS electronics. Recently the use of colloidal quantum dot photodetectors have addressed the CMOS integration challenge for infrared detectors challenge to a great extent,[1-4] yet the presence of Pb may impose some regulatory concerns. In the last years, there is a growing need for the development of environmentally friendly materials, eliminating totally or partially the use of toxic elements such as Pb-, Cd- and Hg-, yet to date application of such kind of semiconductors have been very scarce, since its bandgap is not suitably placed in the infrared.

Benefiting from potential interesting features such as a favourable band gap, high absorption coefficient and environmentally friendly elements, bismuth sulphide ($Bi_2S_3$) has recently attracted attention as an emerging functional material for various applications such as thermoelectric, photodetectors and solar cells.[5-7] The crystal structure of stoichiometric $Bi_2S_3$ is orthorhombic, with a layered structure form of atomic scale ribbons which are held together by van der Waals forces. The structure of $Bi_2S_3$ favours the formation of one-dimensional nanostructures such as nano-wires, rods, tubes and ribbons.[8-11] Two-dimensional (2D) nanosheets and colloidal quantum dots (CQDs) have also been developed.[12-14] $Bi_2S_3$ based photodetectors have been reported yet their performance thus far has not proved to be competitive over other currently available technologies and seems to be limited within the spectral coverage of $Bi_2S_3$ as dictated by its bandgap of ~1.3 eV.[13,15-18] In this work we exploit the formation of mid-gap levels in $Bi_2S_3$ upon formation of sulphur vacancies in reporting photoresponse below the bandgap extended into the short-wave infrared. At the same time we employ a gated photoconductive architecture that reaches very high photoconductive gain, which taken together with the very low noise yields specific detectivities D* close to $10^{15}$ Jones in the visible-near infrared and $10^{12}$ Jones in the short infrared, competing therefore with existing technologies that rely on above bandgap absorption.

## 2. Results and Discussion

### 2.1. Electronic structure

$Bi_2S_3$, the crystal structure of which is shown in the inset of **Figure 1**a, is a binary semiconductor with reported experimental bandgap ranging from 1.2 eV to 1.5 eV.[18-21] Figure 1a illustrates the calculated density of states (DOS) plots obtained at Perdew-Burke-Ernzerhof (PBE) level for bulk $Bi_2S_3$.[22] According to the computational results, S atoms seem to contribute strongly to the formation of the valence band, while Bi atoms have more weight in the corresponding conduction band. However, the formation of S vacancies gives rise to an additional Bi-S interaction within the bandgap of the stoichiometric semiconductor shown in Figure 1b. Such band is calculated to be present below the Fermi level and, therefore, is filled with electrons. This new band essentially acts as a new valence band for the sulphur deficient (SD)-$Bi_2S_3$ and therefore enables the material to be adsorbing in the short-wave infrared through optical transitions. From a theoretical point of view, the origin of the new valence band is related with a consequence produced by the formation of S vacancies, the reduction of $Bi^{3+}$ cations to $Bi^{2+}$. Upon formation of one S vacancy, two electrons are maintained in the system. Calculations show that the most prone element to acquire such electrons are Bi atoms, which are reduced in the process. Essentially, two $Bi^{2+}$ cations are formed per S vacancy. Figure 1c, d illustrate the band structures for both stoichiometric and SD-$Bi_2S_3$, respectively. Indirect-type of band gap is determined for the stoichiometric system, with a band gap value of around 1.38 eV. In the SD case, formation of new Bi-S interactions are present in the form of bands located in the inter band region below the Fermi level, giving rise to a band gap of ~0.79 eV which is reduced compared to that in stoichiometric $Bi_2S_3$ because of the formation of in-gap states enabling the photon absorption in the short-wave infrared range.

**2.2. Materials characterization**

To experimentally approach these conditions we have taken single crystalline flakes of $Bi_2S_3$ (see Experimental section). The Raman spectra (**Figure 2**a) shows the typical and sharp phonon modes of $Bi_2S_3$, which matches well with previous reports.[23] The selected-area electron diffraction (SAED) pattern shows a clear and regular spot pattern (inset of Figure 2a), further confirming the high quality of the single-crystalline feature of the $Bi_2S_3$ flakes. The scanning transmission electron microscopy (STEM) imaging of the single crystal was also measured (Figure 2b) showing a clear orthorhombic atomic structures consisting of weakly interacting one-dimensional ribbons made by tightly bonded [$Bi_4S_6$] units with expected lattice parameters (a = 1.076 nm, b = 0.388 nm, c = 1.065 nm). The red and green balls represent bismuth and sulphur atoms, respectively, only Bi atoms can be seen in Z contrast because of its much heavy mass (Z=83) compared to sulphur (Z=16). As exfoliated pristine $Bi_2S_3$ flakes were found to possess sulphur vacancies. Figure 2c shows the X-ray photoelectron spectroscopy (XPS) of pristine $Bi_2S_3$ flakes, in which Bi 4f doublets and S 2p doublet inside the Bi 4f doublet gap are observed. The main peaks located at 158.4 eV and 163.7 eV are attributed to Bi $4f_{7/2}$ and Bi $4f_{5/2}$ of $Bi^{3+}$ states, respectively.[24] The accompanying weak Bi 4f doublet at lower energy (157.1 eV and 162.4 eV for Bi $4f_{7/2}$ and Bi $4f_{5/2}$, respectively) are assigned to under-coordinated and reduced Bi species with lower valency, $Bi^{2+}$ according to previous reports,[25-27] suggesting the existence of sulphur vacancies in pristine samples. Energy-dispersive X-ray spectroscopy (EDX) (Figure S1a) also shows the nonstoichiometric characteristics of this crystal with atomic percentages of 48.6% and 52% for Bi and S, respectively. In order to synthesize a stoichiometric $Bi_2S_3$ crystal we performed a sulfurization process (see Experimental section). After sulfurization, the $Bi_2S_3$ flakes become stoichiometric with atomic percentages of 39% and 61% for Bi and S, respectively, according to EDX spectra (Figure S1b). Concomitantly, the Bi 4f doublet in XPS spectra shown in Figure 2d at lower binding energies (157.1 eV and 162.4 eV for Bi $4f_{7/2}$ and Bi $4f_{5/2}$, respectively) which are assigned to reduced and under-coordinated Bi species also disappear, suggesting a refilling of the existent sulphur vacancies.

## 2.3. Photodetector performances

*2.3.1. Device scheme and responsivity*

Next, we fabricated photoconductive detectors based on $Bi_2S_3$ flakes on a $Si/SiO_2$ substrate that was used as a backgate to control the conductivity of the $Bi_2S_3$ channel. **Figure 3**a shows a scheme of the photodetector and Figure 3b shows a scanning electron microscope (SEM) image of the detector. The typical thickness of the $Bi_2S_3$ channel was approximately 58 nm, as measured by atomic force microscopy (AFM) shown in Figure 3c. We have considered the two cases of $Bi_2S_3$ photodetectors as mentioned above: the stoichiometric (after sulfurization) and the SD-$Bi_2S_3$ one. From transfer characteristics (Figure S2), the SD-devices exhibit n-type behaviour implying the majority of electrons with density of $10^{17}$ cm$^{-3}$ which can be ascribed to the presence of sulphur vacancies,[14,19] whereas the electrons density drops to be ~$2.7 \times 10^{16}$ cm$^{-3}$ after sulfurization due of the refilling of the vacancies, which is consistent with XPS and EDX measurements. In agreement with DF calculations the stoichiometric $Bi_2S_3$ exhibits photocurrent as determined by its bandgap with an onset at ~1000 nm (Figure 3d). The SD-$Bi_2S_3$ photodetector, on the other hand, exhibits significant photoresponse that extends beyond its bandgap down to 1500 nm, in accordance with the new bandgap states shown in Figure 1b, d. The stoichiometric $Bi_2S_3$ photodetector has responsivity on the order of 600 A/W in the visible (Figure 3d) exhibiting a nonlinear response which reaches values up to $5 \times 10^4$ A/W at low light intensities on the order of nW/cm$^2$ (Figure 3e). The SD photodetector exhibits responsivity close to $10^4$ A/W in the visible and 10 A/W in the infrared at 1500 nm (Figure 3d). At low illumination intensity the SD-photodetector reaches very high values of responsivity exceeding $10^7$ A/W at 0.1 nW/cm$^2$ (Figure 3e). All the values were achieved at drain bias of 1 V and zero gate voltage.

The effect of gate bias on the photocurrent and responsivity of both detectors is illustrated in **Figure 4**. **Figure 4a** shows the responsivities of SD-devices, as a function of irradiance, under different gate voltage ($V_g$). At high irradiance values the responsivity of the detector is independent of the gate bias, while at low irradiance exhibits higher responsivity at positive back gate values (channel accumulation regime) than in negative back gate values (channel depletion regime). We attribute this to the effective contact barrier in depletion mode at negative gate (bottom inset in Figure 4a) which suppresses carrier recirculation. Under positive bias and illumination, the electron quasi-Fermi level is close to the conduction band leading to efficient carrier injection/extraction at the metal/Bi$_2$S$_3$ interfaces of the source and drain electrodes facilitating efficient recirculation (top inset in Figure 4a). Figure 4b shows the responsivity of stoichiometric devices (i.e. upon sulfurization), also exhibiting the same dependence upon $V_g$. In both detectors, the responsivity shows no appreciable dependence on gate voltage at high incident light power, which confirms that the underlying photodetection mechanism is the photoconductive effect.[28]

*2.3.2. Temporal response*

The time response of the two photodetectors is compared in **Figure 5**a, b for visible excitation. The stoichiometric detector exhibits fast photoresponse with response time on the order of 5 ms (Figure 5b). The SD-detector is characterized by a response time on the order of 500 ms (Figure 5a), approximately two orders of magnitude longer than the stoichiometric one. This fact is in agreement with the corresponding improvement of responsivity of two orders of magnitude on the SD-detector compared with the stoichiometric one and underpins the origin of high responsivity of the SD-detector. The gate voltage also does not have influnce on the speed of both devices (Figure S3). However, for most practical applications, photodetector speed is an issue and it is desirable to have response times on the order of ms or faster, for sensing or imaging applications. We therefore sought to improve the response speed of the SD-based detector without sacrificing its spectral coverage into the infrared. We found that a mild chemically treatment (see Experimental section) on the SD-$Bi_2S_3$ leads to a significant acceleration of the photodetector speed with a time response on the order of 10 ms for infrared (Figure 5c) and visible light excitation (Figure S4). This acceleration was accompanied by a 10-fold reduction in responsivity that reached values over $10^6$ A/W at intensities on the order of nW/cm$^2$ (Figure 3e). Most favourably the mild treatment preserved the SD-detectors' infrared response with responsivities up to 10 A/W measured at 1600 nm (Figure 3d).

*2.3.3. Noise and sensitivity*

The most relevant figure of merit that characterizes the sensitivity of photodetectors is the specific detectivity D* defined as $D^* = \frac{\sqrt{AB}}{NEP} = \frac{R\sqrt{A}}{S_n}$, where NEP is the noise equivalent power, *R* is the responsivity, *A* is the active area of the detector, *B* is the electrical bandwidth, and $S_n$ is the noise spectral density of the detector. In order to report the D* of this set of detectors we have performed noise measurements as shown in **Figure 6**a. The noise spectral density ($S_n$) was obtained by calculating the Fourier transformation of dark current traces at a sampling rate of 50 Hz under exactly the same conditions as the optical measurements were performed. In all cases, a 1/*f*-noise component is observed due to the charge trap states in $Bi_2S_3$ channel or the interface between $Bi_2S_3$ and substrate, similar to the case of graphene and TMDs based detectors.[29,30] All devices exhibit a very small noise density ranging from 1 pA Hz$^{-1/2}$ to 30 pA Hz$^{-1/2}$ at frequency of 1 Hz. Considering the fast speed of the detectors on the order of 10 ms after sulfurization and chemically treatment, the noise spectral density can be extracted at higher frequency (10 Hz) for calculating the specific D* in these devices.

Considering together the noise current, speed and responsivity at 1 nW/cm$^2$ of those detectors, we plot in Figure 6b the spectral D* of the three detectors. As evidenced the SD-detectors reach a D* as high as ~4×10$^{14}$ Jones in the visible and 5 × 10$^{11}$ Jones at 1.5 µm at bandwidth of 1 Hz and external bias of 1 V due to the high responsivity and low noise. After sulfurization, the devices have a drop in D* to 10$^{13}$ Jones in visible and limited spectral coverage to 1.1 µm. It is noted that the SD-detectors after chemically treatment still hold a high D* close to 10$^{15}$ Jones in the visible-near infrared range and 10$^{12}$ Jones at 1.6 µm at 10 Hz benefiting from the faster speed and high responsivity.

## 3. Conclusion

In summary, we have reported a highly sensitive CMOS compatible infrared photodetector based on vacancy-engineered Bi$_2$S$_3$ semiconductor exploiting the formation of extended in-gap states that allow for optical absorption below the bandgap value of Bi$_2$S$_3$. We have made use of this knowledge to build a photoconductive photodetector with high gain in order to develop a high sensitivity photodetector. This work illustrates that defects in semiconductors may be used favourably to deliver high performance optoelectronic devices with functionalities that cannot be reached by semiconductor materials in its defect-free state. This work also expands the material availability for infrared optoelectronics.

## 4. Experimental Section

*Device fabrication and characterization:* The bulk Bi$_2$S$_3$ crystals were purchased from the 2D semiconductors corporation. The thin Bi$_2$S$_3$ flakes was then exfoliated with polydimethylsiloxane (PDMS) tape on Si/SiO$_2$ (285 nm) wafer using the micromechanical exfoliation method. The bulk Bi$_2$S$_3$ slices were pasted on PDMS adhesive tape. The sticky side of the tape was folded in half and then torn slowly to peel off the crystals. Repeat the above operation for several times and finally exfoliate the thin flakes onto the substrate by pressing the tape gently and removing it slowly. Before the device fabrication, the substrate with Bi$_2$S$_3$ flakes was soaked into acetone for 2 h at 60 °C to remove the residual glue. Metal contacts were then fabricated by the laser writing lithography, and Ti (2 nm) and Au (70 nm) electrodes were evaporated by e-beam and thermal evaporation, respectively.

All the measurements were performed in ambient conditions using an Agilent B1500A semiconducting device analyzer. Responsivity and temporal response times were measured under pulsed light at a wavelength of 635 nm from a four-channel laser controlled with an Agilent A33220A waveform generator. The spectral photo-response measurements were performed with fiber-coupled light from a supercontinuum light source (SuperKExtreme EXW-4, NKT Photonics). STEM experiments were performed in a probe corrected FEI Titan 60-300 microscope operated at 300 kV. This microscope is equipped with a high-brightness field emission gun (X-FEG) and a CETCOR corrector for the condenser system, which produces an electron probe with a lateral size below 1 Å. Z contrast STEM imaging was carried out with a Fischione 3000 annular dark field detector. Chemical analysis was performed in this microscope by Energy dispersive X-rays spectroscopy (EDX) in an EDAX detector equipped with Genesis RTEM software embedded.

*Sulfurization and chemically treatment:* The sulfurization was performed in horizontal tube furnace. The substrates with Bi$_2$S$_3$ flakes based devices were located on the center of the quartz tube, the ceramic boat containing ~200 mg sulfur powers was located on upstream side with a distance of ~20 cm from centre. The tube was pumped down to $2\times10^{-2}$ mbar to remove the air fully in the tube, then refill the tube with ultrahigh purity N$_2$ gas. The furnace was heated up to 450 °C in 15 min with N$_2$ flow of around 10 sccm, then keep that temperature for 20 min, afterwards cool it to room temperature naturally.

For the mild chemically treatment, firstly 20 μL bis(trimethylsilyl)-selenide (TMSe) (stored in glove box) was mixed with 2 mL methanol in a small vial. The mixed solution (1% TMSe) was sonicated for 10 min. The substrates with Bi$_2$S$_3$ flakes based devices were soaked into the solution for 3 min and then anneal it on a hotplate with temperature of 100 °C and annealing time of 10 min. The devices were measured before treatment and after treatment immediately.

*DF calculation:* Calculations have been performed using the Vienna *ab initio* package VASP.[31] The exchange-correlation functional chosen for the study was the Perdew-Burke-Ernzerhof (PBE),[22] a widely used functional from the generalized gradient approximation (GGA). Since VASP uses plane-waves basis to represent the electronic density, 415 eV were used as cut-off kinetic energy. The interaction of core electrons with the valence electronic density was modelled using the projected augmented-wave (PAW).[32,33] A Monkhorst-Pack mesh of 3x3x3 k-points was used for unit cell.[34] Atoms and cell were fully relaxed until forces acting on atoms were below 0.01 eV Å$^{-1}$. In order to take into account the possible electron transfer produced in defective systems, spin-polarized calculations were performed. A Gaussian smearing with an energy window of 0.1 eV was used, even though final energies were extrapolated to 0 K. Bader analysis and magnetization values were used in order to assign atomic charges.[35,36] Defective $Bi_2S_3$ has been modelled by removing 8% of S atoms present in the lattice of the supercell.

**Supporting Information**

Supporting Information is available from the Wiley Online Library or from the author.

**Acknowledgements**

N.H. and A.F. contribute equally. The authors acknowledge financial support from the European Research Council (ERC) under the European Union's Horizon 2020 research and innovation programme (grant agreement no. 725165), the Spanish Ministry of Economy and Competitiveness (MINECO), and the "Fondo Europeo de Desarrollo Regional" (FEDER) through grant TEC2017-88655-R. The authors also acknowledge financial support from Fundacio Privada Cellex, the program CERCA and from the Spanish Ministry of Economy and Competitiveness, through the "Severo Ochoa" Programme for Centres of Excellence in R&D (SEV-2015-0522). Y.Y. acknowledges support from the National Natural Science Foundation of China under Grant NO.61805045. S.C. acknowledges support from a Marie Curie Standard European Fellowship ("NAROBAND", H2020-MSCA-IF-2016-750600).

Furthermore, the research leading to these results has received funding from the European Union H2020 Programme under grant agreement n°696656 Graphene Flagship.

Received: ((will be filled in by the editorial staff))
Revised: ((will be filled in by the editorial staff))
Published online: ((will be filled in by the editorial staff))

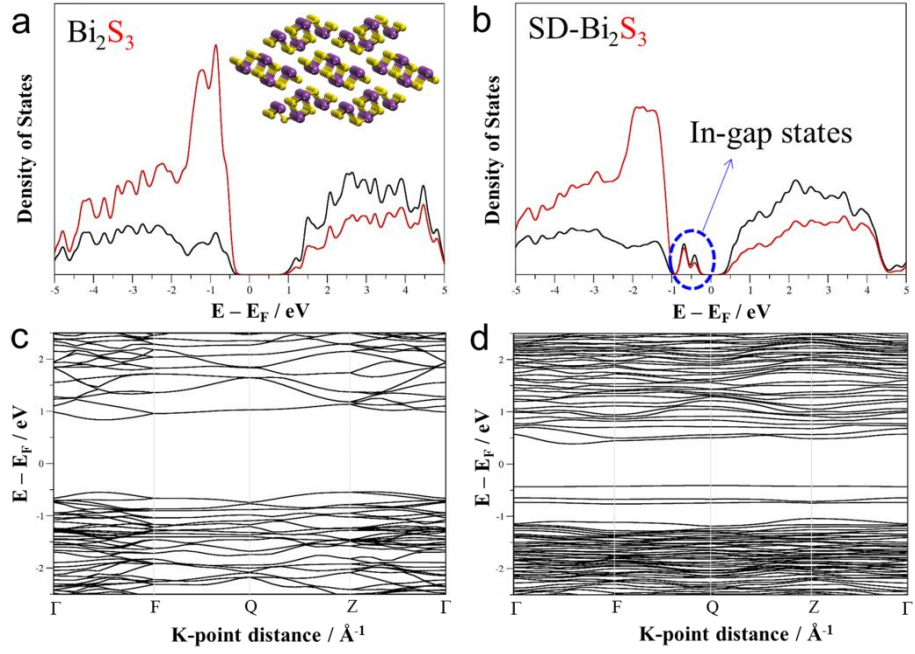

**Figure 1.** (a) Calculated density of states (DOS) and crystal structure of stoichiometric $Bi_2S_3$. Black and red lines correspond to Bi and S, respectively. For the crystal structure model, purple balls represent Bi atoms, while yellow balls represent S ones. (b) DOS plot of S-deficient (SD)-$Bi_2S_3$. Band structures of (c) stoichiometric $Bi_2S_3$ and (d) SD-$Bi_2S_3$.

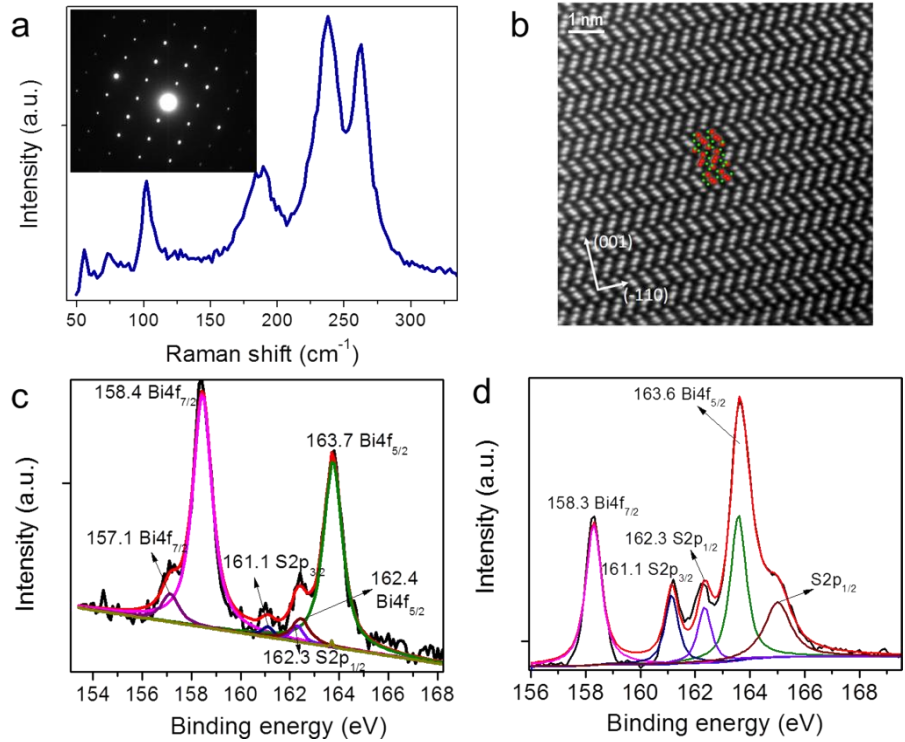

**Figure 2.** (a) Raman spectra of the $Bi_2S_3$ flakes, the inset is corresponding SAED pattern. (b) Z contrast STEM imaging of the single crystal along (100) zone axis. The red and green balls represent bismuth and sulphur atoms, respectively. (c) XPS of SD-$Bi_2S_3$, showing the weak Bi 4f doublet at lower binding energy (157.1 eV and 162.4 eV for Bi $4f_{7/2}$ and Bi $4f_{5/2}$, respectively) which is attributed to the under-coordinate and reduced Bi species. (d) XPS of stoichiometric $Bi_2S_3$ after sulfurization.

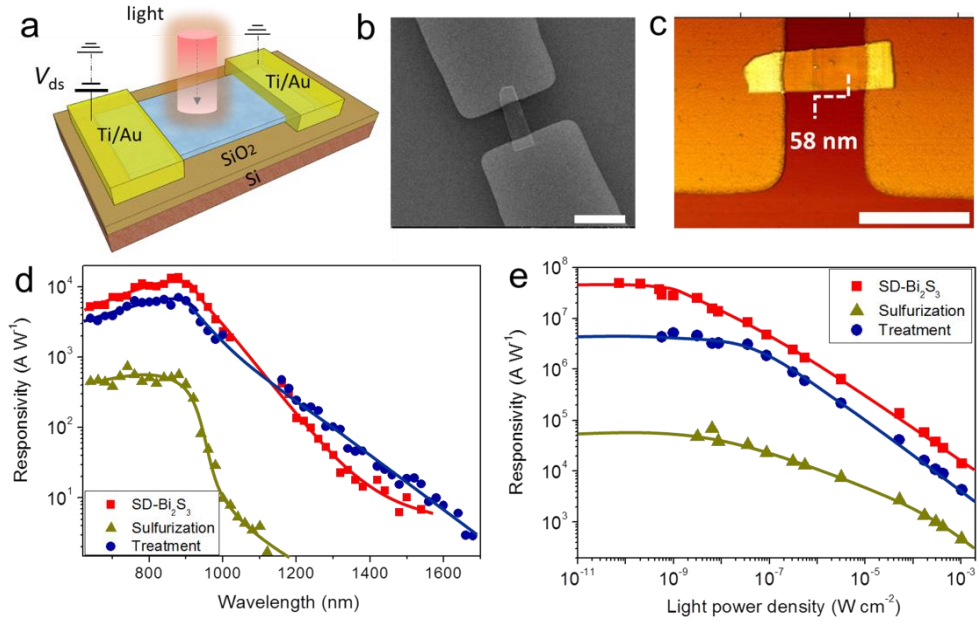

**Figure 3.** (a) Schematic diagram of $Bi_2S_3$ detectors on $Si/SiO_2$ substrate. (b) Scanning electron microscope (SEM) image and (c) atomic force microscopy (AFM) image of the typical $Bi_2S_3$ device. Both scale bars are 5 μm. (d) Spectral responsivities for SD-$Bi_2S_3$, stoichiometric and chemically treated devices. (e) Responsivities of the different set of detectors as a function of incident light power density with wavelength of 635 nm. The solid lines are the guides to eye.

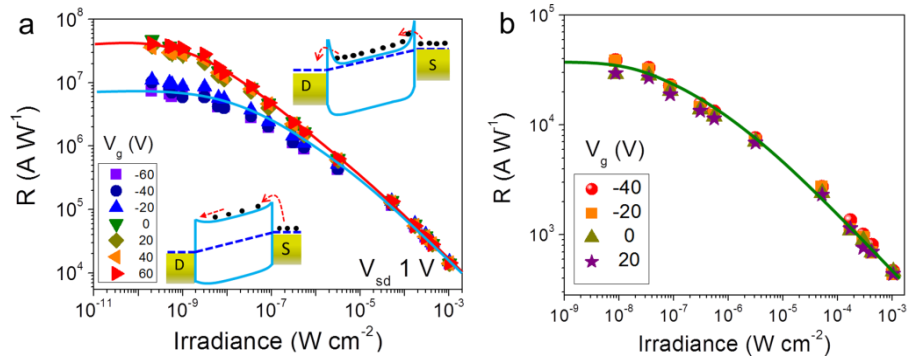

**Figure 4.** (a) Responsivities of the SD-device as a function of irradiance at different $V_g$ and $V_{sd}$ of 1 V. Top inset is the band diagram of the device at positive gate under high light illumination, showing the sharp band bending at $Bi_2S_3$ and Au interface. Bottom inset is the band diagram at negative gate voltage and under low irradiance. (b) Responsivities of the stoichiometric device as a function of irradiance at different $V_g$ and $V_{sd}$ of 1 V.

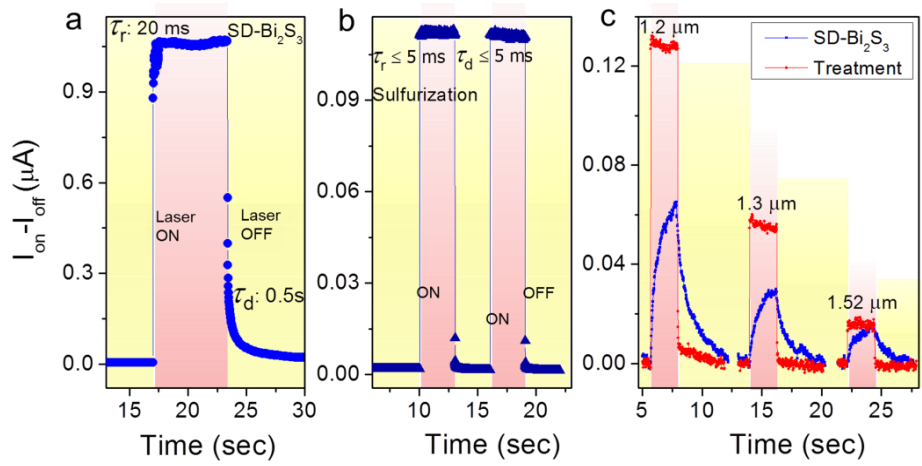

**Figure 5.** Temporal response of (a) the SD-Bi$_2$S$_3$ detector and (b) the stoichiometric device under light illumination with wavelength of 635 nm and power density of 1 mW cm$^{-2}$. (c) Temporal response of SD-Bi$_2$S$_3$ detectors before and after chemically treatment under infrared (1.2-1.5 µm) light illumination.

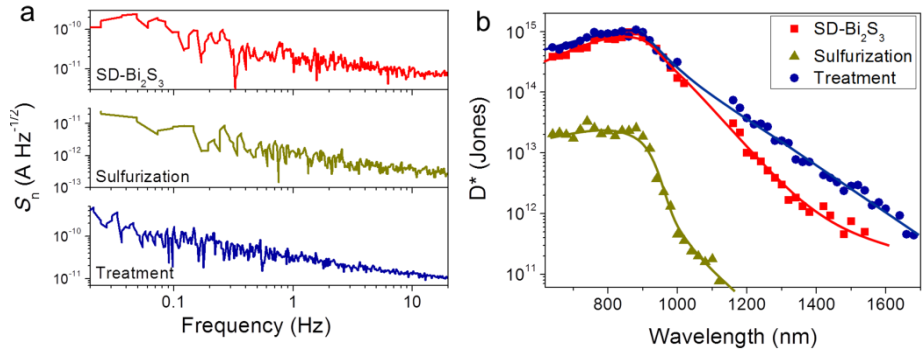

**Figure 6.** (a) Noise spectral density ($S_n$) for all classes of $Bi_2S_3$ detectors. (b) Spectral D* for all devices by extracting the responsivity at low irradiance and the noise spectral density at bandwidth of 1 Hz for SD-detectors and 10 Hz for sulfurized and treated devices. The solid lines are the guides to eye.

# Supporting Information

**Engineering Vacancies in Bi$_2$S$_3$ yields sub-Bandgap Photoresponse and highly sensitive Short-Wave Infrared Photodetectors**

*Nengjie Huo, Alberto Figueroba, Yujue Yang, Sotirios Christodoulou, Alexandros Stavrinadis, César Magén, Gerasimos Konstantatos\**

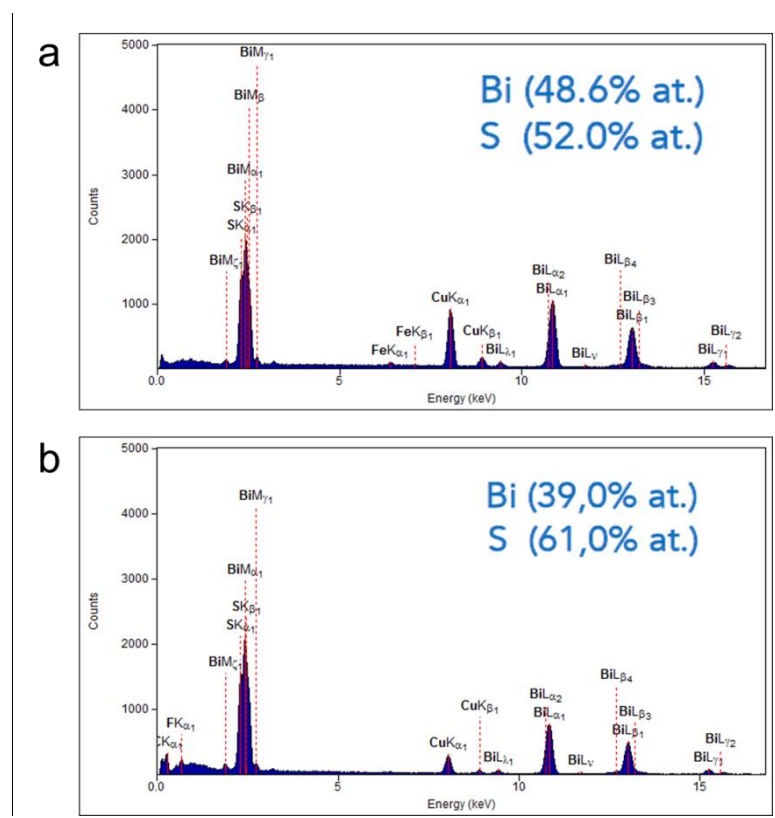

Figure S1. (a) Energy-dispersive X-ray spectroscopy (EDX) spectral of pristine Bi$_2$S$_3$ single crystal, showing the substoichiometric feature with sulphur vacancies. (b) EDX spectral of Bi$_2$S$_3$ flakes after sulfurization, the samples become stoichiometric, indicating the refilling of the sulphur vacancies.

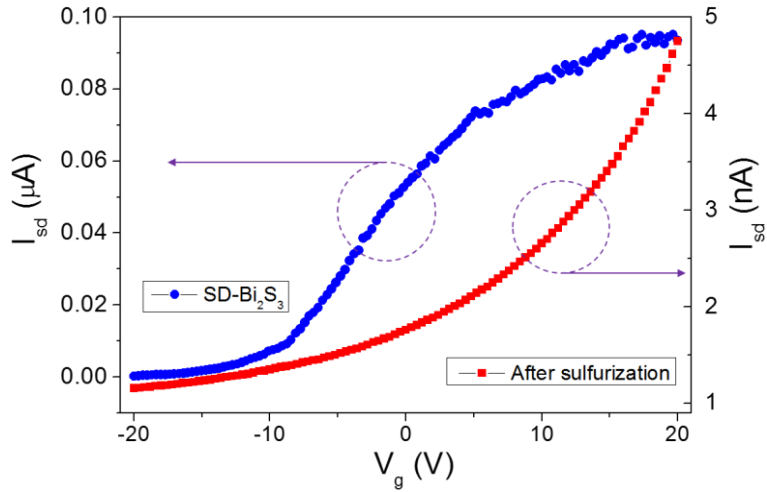

Figure S2. Transfer characteristics of SD-Bi$_2$S$_3$ based detectors before and after sulfurization, showing the n-type behaviour because of the existence of sulphur vacancies.

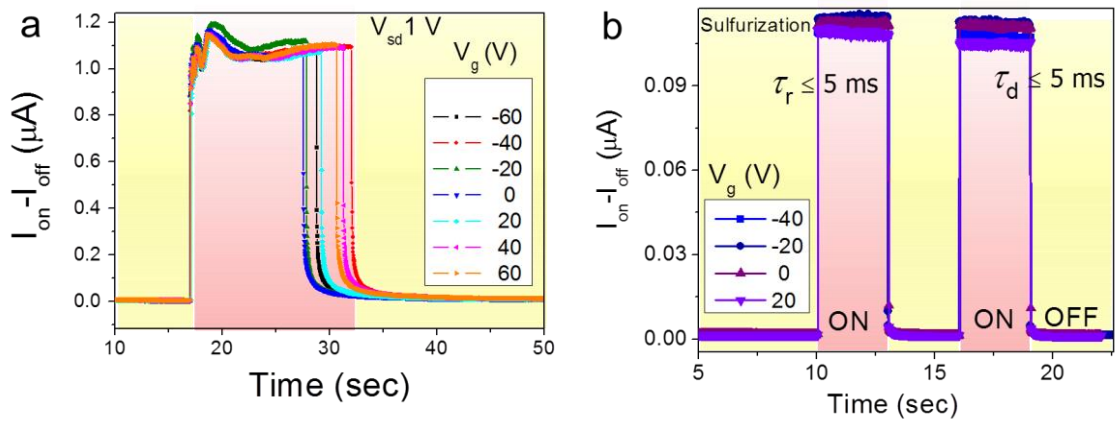

Figure S3. (a) Dynamic photoresponse of SD-Bi$_2$S$_3$ detectors at V$_{sd}$ of 1 V and varying back gate voltage (V$_g$) with 635 nm light illumination. The photocurrent and response speed keep almost same under different V$_g$. (b) Dynamic photoresponse of stoichiometric Bi$_2$S$_3$ detectors after sulfurization, showing the faster speed and unchanged photocurrent with varying back gate voltage.

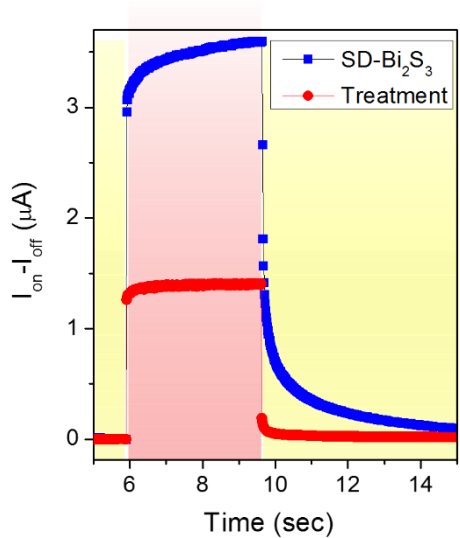

Figure S4. Dynamic photoresponse of SD-$Bi_2S_3$ detectors before and after chemically treatment under visible light illumination with wavelength of 635 nm. The speed has been much accelerated to be on the order of 10 ms under visible light excitation after chemically treatment.